\documentclass[11pt,a4paper]{article}
\pdfoutput=1
\usepackage{jcappub}

\newcommand{\be}{\begin{equation}}
\newcommand{\ee}{\end{equation}}
\newcommand{\bea}{\begin{eqnarray}}
\newcommand{\eea}{\end{eqnarray}}
\newcommand\pp{\,\,\,.}
\newcommand\vv{\,\,\,,}

\subheader{\hfill{\tt IFIC/10-38}}

\title{Constraining the cosmic radiation density due to lepton number with Big Bang Nucleosynthesis}

\author[a]{Gianpiero Mangano,}
\author[a,b]{Gennaro Miele,}
\author[c]{Sergio Pastor,}
\author[a,b]{Ofelia Pisanti,}
\author[a,b]{and Srdjan Sarikas}

\affiliation[a] {Istituto Nazionale di Fisica
 Nucleare - Sezione di Napoli \\ Complesso Universitario di Monte
 S.~Angelo, I-80126 Napoli, Italy}
\affiliation[b] {Dipartimento di Scienze Fisiche, Universit\`{a} di Napoli
{\it Federico II} \\ Complesso Universitario di Monte S.~Angelo, I-80126
Napoli, Italy} \affiliation[c] {Instituto de F\'{\i}sica Corpuscular
(CSIC-Universitat de Val\`{e}ncia),\\ Ed.\ Institutos de
Investigaci\'{o}n, Apdo.\ correos 22085, E-46071 Valencia, Spain}

\emailAdd{mangano@na.infn.it}
\emailAdd{miele@na.infn.it}
\emailAdd{pastor@ific.uv.es}
\emailAdd{pisanti@na.infn.it}
\emailAdd{sarikas@na.infn.it}

\abstract{ The cosmic energy density in the form of radiation before and during
Big Bang Nucleosynthesis (BBN) is typically parameterized in terms of the
effective number of neutrinos $N_{\rm eff}$. This quantity, in
case of no extra degrees of freedom, depends upon the chemical potential and the temperature characterizing
the three active neutrino distributions, as well as by their possible non-thermal features. In
the present analysis we determine the upper bounds that BBN places on $N_{\rm eff}$ from 
primordial neutrino--antineutrino asymmetries, with a careful treatment of the dynamics of neutrino oscillations. We consider quite
a wide range for the  total lepton number in the neutrino sector,
$\eta_\nu= \eta_{\nu_e}+\eta_{\nu_\mu}+\eta_{\nu_\tau}$ and the initial
electron neutrino asymmetry $\eta_{\nu_e}^{\rm in}$, solving the
corresponding kinetic equations which rule the dynamics of neutrino
(antineutrino) distributions in phase space due to collisions, pair
processes and flavor oscillations. New bounds on both the total lepton
number in the neutrino sector and the $\nu_e -\bar{\nu}_e$ asymmetry at
the onset of BBN are obtained fully exploiting the time evolution of
neutrino distributions, as well as the most recent determinations of
primordial $^2$H/H density ratio and $^4$He mass fraction. Note that
taking the baryon fraction as measured by WMAP, the  $^2$H/H abundance plays a
relevant role in constraining the allowed regions in the $\eta_\nu - \eta_{\nu_e}^{\rm in}$  plane.
These bounds fix the maximum contribution of neutrinos with
primordial asymmetries to $N_{\rm eff}$ as a function of the mixing
parameter $\theta_{13}$, and point out the upper bound $N_{\rm eff}
\lesssim 3.4$. Comparing these results with the forthcoming
measurement of $N_{\rm eff}$ by  the Planck satellite will likely provide
insight on the nature of the radiation content of the universe. }

\keywords{Neutrinos, physics of the early universe, primordial asymmetries}
\arxivnumber{1011.0916}

\notoc
\begin{document}
\maketitle

%%%%%%%%%%%%%%%%%%%%%%%%%%%%%%%%%%%%%%%%%%%%%%%%%%%%%%%%%%%%%%%%%%%%%%
\section{Introduction}\label{sec:introduction}
%%%%%%%%%%%%%%%%%%%%%%%%%%%%%%%%%%%%%%%%%%%%%%%%%%%%%%%%%%%%%%%%%%%%%%

The dynamics of neutrino oscillations in the early universe has been extensively
studied in the scientific literature and  is expected to have produced an efficient
mixing of flavor neutrino distributions at a temperature $T_\gamma \sim 1$ MeV,
thus shortly before the freezing of the neutron-to-proton density ratio
and the onset of Big Bang Nucleosynthesis (BBN). Indeed, while flavor
neutrino conversions are suppressed by matter effects at larger
temperatures, at $T_\gamma \sim 10$ MeV the atmospheric mass-squared
difference $\Delta m^2_{\rm atm}$ triggers efficient $\nu_\mu-\nu_\tau$ mixing, as well as
$\nu_x-\nu_e$ ($x=\mu,\tau$) conversions if the angle $\theta_{13}$ is not vanishing
and sufficiently large. Finally, at $T_\gamma\lesssim 3$ MeV oscillations
driven by  $\Delta m^2_{\rm sol}$ set on and are large enough to achieve
strong flavor conversions before BBN
\cite{Lunardini:2000fy,Dolgov:2002ab,Wong:2002fa,Abazajian:2002qx}.

These results have a major impact on the possible values for the
cosmological lepton asymmetry stored in each neutrino flavor, which in
analogy with the baryon-antibaryon asymmetry parameter
$\eta_b=(n_b-n_{\bar{b}})/n_\gamma$, can be parameterized by the number
density ratios
\be
\eta_{\nu_\alpha} = \frac{n_{\nu_\alpha}-n_{\bar{\nu}_\alpha}}{n_\gamma}\vv \,\,\, \alpha=e,\mu,\tau \pp
\label{etanualpha}
\ee
Based on the equilibration of lepton and baryon asymmetries by sphalerons
in the very early universe, such a neutrino asymmetry should be of the
same order of the cosmological  baryon number $\eta_b=273.93 \,\Omega_b
h^2\,  10^{-10}$, which is restricted to be a few times $10^{-10}$ by
present observations, such as 7-year data from the WMAP satellite and
other cosmological measurements \cite{Komatsu:2010fb}. Nevertheless, a cosmological
neutrino asymmetry  orders of magnitude larger than this value is still an
open possibility, with implications on fundamental physics in the early
universe, such as their potential relation with the cosmological
magnetic fields at large scales \cite{Semikoz:2009ye}.  
In particular, a non-zero lepton asymmetry leads to an enhanced
contribution of neutrinos to the energy density in the
form of radiation $\rho_r$, which, after the $e^+e^-$ annihilation phase, 
is usually parameterized as
$\rho_r/\rho_\gamma=1 + 7/8 (4/11)^{4/3} N_{\rm eff}$. The parameter 
$N_{\rm{eff}}$ is the ``effective number of neutrinos" whose standard
value is 3 in the limit of instantaneous neutrino decoupling. 
Interestingly, recent data on the anisotropies of the cosmic
microwave background from WMAP \cite{Komatsu:2010fb} and the primordial
$^4$He abundance \cite{Izotov:2010ca,Aver:2010wq,Aver:2010wd} 
seem to favor a value of $N_{\rm{eff}}>3$,
although with large errorbars.

If neutrinos are in kinetic and chemical equilibrium, their distribution
of momenta is parameterized by a temperature and a well defined chemical
potential $\mu_{\nu_\alpha}$, and each flavor neutrino asymmetry can be
expressed in terms of the corresponding degeneracy parameter
$\xi_{\nu_\alpha}\equiv \mu_{\nu_\alpha}/T_{\nu_\alpha}$ as
\be
\eta_{\nu_\alpha} = \frac{1}{12 \zeta(3)}
\left(\frac{T_{\nu_\alpha}}{T_\gamma}\right)^3 \left( \pi^2
\xi_{\nu_\alpha}+\xi_{\nu_\alpha}^3 \right) \pp
\label{eta-eq}
\ee
In this case, if flavor oscillations enforce the condition that all
$\xi_{\nu_\alpha}$ are almost the same during BBN, the stringent bound on
the electron neutrino degeneracy, which directly enters the neutron/proton
chemical equilibrium \cite{KangSteigman}, applies to all flavors. The
common value of the neutrino degeneracies is restricted to the range
$-0.021 \leq \xi_{\nu_\alpha} \leq 0.005$, $ \alpha =e, \mu, \tau$
\cite{Iocco:2008va} (see also
\cite{Hansen:2001hi,Barger:2003zg,Barger:2003rt,Cuoco:2003cu,Cyburt:2004yc,
Serpico:2005bc,Simha:2008mt,Krauss:2010xg} for other analyses). This in
turn also implies that, \emph{if neutrinos indeed, reach perfect kinetic
and chemical equilibrium before they decouple}, any large excess  in
cosmic radiation density, if observed, must be ascribed to extra
relativistic degrees of freedom since the additional contribution to
radiation density due to non vanishing $\xi_{\nu_\alpha}$ is very
small. 
Assuming
that the neutrino distributions are given by their equilibrium form, the
BBN bound on the neutrino degeneracies leads to the following upper limit on the
excess contribution to $N_{\rm{eff}}$ \cite{Dolgov:2002ab},
\be
\Delta N_{\rm{eff}} = \sum_{\alpha=e,\mu,\tau}
\left[\frac{30}{7}\left( \frac{\xi_{\nu_\alpha}}{\pi} \right)^2 +
\frac{15}{7}\left( \frac{\xi_{\nu_\alpha}}{\pi} \right)^4 \right] \lesssim
0.0006 \vv \label{deltan}
\ee
which is tiny, even compared with the value of $N_{\rm eff}=3.046$ found
solving the neutrino kinetic equations in absence of asymmetries
\cite{Mangano:2005cc}.

Neutrino kinetic and chemical equilibrium is maintained  in the early
universe by purely leptonic weak processes such as neutrino-neutrino
interactions, $\nu$-$e^\pm$ scatterings and pair processes, $\nu
\bar{\nu}\leftrightarrow e^+e^-$, whose rates become of the order of the
Hubble parameter at $T_\gamma \sim 2-3$ MeV. Baryons play no role in this
concern due to their very low density. At lower temperatures weak
interactions are no longer effective, neutrinos decouple from the rest of
the primeval plasma and their distribution in phase space is frozen out.
Flavor oscillations driven by $\Delta m^2_{\rm atm}$ take place when
neutrinos are still fastly scattering off the surrounding medium, so that
the changes in their distribution due to oscillations are efficiently
readjusted into an equilibrium Fermi-Dirac function. Instead, flavor
conversions due to $\Delta m^2_{\rm sol}$ and $\theta_{12}$ occur around
neutrino decoupling. This implies, at least in principle, that if
neutrinos succeed in achieving comparable asymmetries in all flavors
before BBN, their distributions might acquire
distortions with respect to equilibrium values due to inefficient
interactions.

This can be easily understood by a simple  example. Suppose that at
temperatures higher than $2-3$ MeV we start with a vanishing total
asymmetry, but  $\eta_{\nu_e}^{\rm in}=- 2 \eta_{\nu_x}^{\rm in} \neq 0$ and we artificially
switch-off scattering and pair processes. Due to solar-scale oscillations,
in the case of maximal mixing asymmetries in each flavor will eventually
vanish, but the neutrino distributions will not correspond to equilibrium,
since averaging two equilibrium distributions with a different chemical
potential does not correspond to a Fermi-Dirac function. Only scatterings
and pair-processes can turn it into an equilibrium distribution with, in
this case, zero chemical potential.

Though this example is quite extreme and unrealistic, nevertheless, it
tells us that the interplay of neutrino freeze--out and $\Delta m^2_{\rm
sol}$ oscillation phases might deserve a more careful scrutiny, as first
discussed in \cite{Pastor:2008ti}. In fact, depending on the initial
flavor neutrino asymmetries and the value of $\theta_{13}$, the final
neutrino distributions at the onset of BBN might show non-thermal
distortions which change the neutron-proton chemical equilibrium due to
the direct role played by electron (anti)neutrinos. Moreover, this
corresponds to an asymmetry-depending parameter $N_{\rm{eff}}>3$ which is
due to inefficient entropy transfer to the electromagnetic plasma and is
not given by the equilibrium value of  eq.\ \eqref{deltan}. As pointed out
in \cite{Pastor:2008ti} these features will prove to be quite important in
case of large initial asymmetries and opposite values for $\nu_e$ and
$\nu_{\mu,\tau}$ chemical potentials. It was shown that with fine-tuned
initial asymmetries the BBN bound could be respected and at the same time
an excess radiation density could survive, corresponding to 
values of $\Delta N_{\rm{eff}}$ of order unity or larger.

In the present work we extend the analysis of \cite{Pastor:2008ti} in two
ways. First, we consider a wider range of values of the initial neutrino
asymmetries and solve their evolution with the corresponding kinetic
equations, including both collisions and oscillations. Moreover, the obtained shape
of the neutrino distributions is then plugged into the BBN dynamics
allowing, by the comparison between the theoretical results and the
experimental data on primordial abundances  of deuterium and $^4$He, to
find more accurate bounds on the total lepton asymmetry stored in the
neutrino sector, as well as the way it was distributed at some early stage
in the $\nu_e$ and $\nu_x$ flavors. In fact, for all initial values of
neutrino asymmetries which are compatible with BBN bounds, the distortions
in the neutrino distribution are typically quite small, see Section
\ref{sec:dynamics}, so that it is accurate enough for our purposes to
parameterize them in terms of a Fermi-Dirac function with two
time-dependent parameters, which correspond to the first two moments of
the actual distribution: an effective chemical potential
$\xi_{\nu_\alpha}$ and an effective temperature $T_{\nu_\alpha}$, or
equivalently, the asymmetry in each flavor and the energy density.

The paper is organized as follows. After setting in Section
\ref{sec:dynamics} the formalism of kinetic equations which rule the
evolution of neutrino distributions written in the standard density matrix
formalism and showing an example of their dynamics, we then study the BBN
constraints on neutrino asymmetries in Section \ref{sec:BBN}. In
particular, we discuss the experimental data which are used in our analysis
for $^2$H/H and $^4$He mass fraction $Y_p$, as well as the way we have
modified the public BBN numerical code $\mathtt{PArthENoPE}$
\cite{Pisanti:2007hk,serpico,parthenope} to track neutrino evolution.
Finally, we report the bounds on initial (at $T_\gamma \sim 10$ MeV)
neutrino asymmetries or their final values after flavor oscillation phase.
In Section \ref{sec:conclusions} we give our concluding remarks.

%%%%%%%%%%%%%%%%%%%%%%%%%%%%%%%%%%%%%%%%%%%%%%%%%%%%%%%%%%%%%%%%%%%%%%
\section{The dynamics of neutrinos in the early universe with primordial asymmetries}\label{sec:dynamics}
%%%%%%%%%%%%%%%%%%%%%%%%%%%%%%%%%%%%%%%%%%%%%%%%%%%%%%%%%%%%%%%%%%%%%%

Our first aim is to calculate the evolution of the three active neutrino distributions in the epoch of the
universe right before BBN, when these particles were interacting among
themselves and with electrons and positrons. The corresponding weak
collision rate decreases very fast with the expansion until neutrinos
decouple at $T_\gamma \sim 1$ MeV. At the same time, it was shown in
\cite{Lunardini:2000fy,Dolgov:2002ab,Wong:2002fa,Abazajian:2002qx,Gava:2010kz}
that for the neutrino mixing parameters currently allowed, flavor
oscillations start to be effective at similar temperatures. In such a case
the best way to describe neutrino distributions  is to use matrices in
flavor space \cite{Sigl:1993fn, McKellar:1994ja}. Since we consider only
the three active neutrino species, we will need $3\times3$ matrices in
flavor space $\varrho_{\bf p}$ for each neutrino momentum ${\bf p}$, where
the diagonal elements are the usual flavor distribution functions
(occupation numbers) and the off-diagonal ones encode phase information
and vanish for zero mixing.

Oscillations in flavor space of the three active neutrinos  are driven by two mass-squared
differences and three mixing angles bounded by the experimental
observations in the following ranges: the "solar" $\Delta m^2_{\rm
sol}=7.59^{+0.44}_{-0.37} \times 10^{-5}$ eV$^2$, the "atmospheric"
$|\Delta m^2_{\rm atm}|= 2.40^{+0.24}_{-0.22} \times 10^{-3}$ eV$^2$ and
correspondingly, the large mixing angles  $\sin^2 \theta_{12}=
0.32^{+0.04}_{-0.03}$ and $\sin^2 \theta_{23}= 0.50^{+0.13}_{-0.11}$
($2\sigma$ ranges from \cite{Schwetz:2008er}). On the other hand,  the third
angle is quite small, $\sin^2 \theta_{13} \leq 0.053$ ($3\sigma)$
\cite{Schwetz:2008er},  even compatible with a vanishing value, though a
mild evidence for $\sin^2 \theta_{13}>0$ has been found in global data
analyses (see e.g.\ \cite{Fogli:2008jx,GonzalezGarcia:2010er}).

The equations of motion (EOMs) for $\varrho_{\bf p}$ are the same as those
considered in reference \cite{Pastor:2008ti},
\begin{equation}
{\rm i}\,\frac{d\varrho_{\bf p}}{dt} =[{\sf\Omega}_{\bf p},\varrho_{\bf
p}]+ C[\varrho_{\bf p},\bar\varrho_{\bf p}]\,,
\label{drhodt}
\end{equation}
and similar for the antineutrino matrices $\bar\varrho_{\bf p}$. The first
term on the r.h.s.\ describes flavor oscillations,
\begin{equation}
{\sf\Omega}_{\bf p}=\frac{{\sf M}^2}{2p}+
\sqrt{2}\,G_{\rm F}\left(-\frac{8p}{3 m_{\rm w}^2}\,{\sf E}
+\varrho-\bar\varrho\right)\,,
\end{equation}
where $p=|{\bf p}|$ and ${\sf M}$ is the neutrino mass matrix (opposite
sign for antineutrinos). Matter effects are included via the term
proportional to the Fermi constant $G_{\rm F}$, where ${\sf E}$ is the
$3\times3$ flavor matrix of charged-lepton energy densities
\cite{Sigl:1993fn}. For our range of temperatures we only need to include
the contribution of electrons and positrons. Finally, the last term arises
from neutrino-neutrino interactions and is proportional to
$\varrho-\bar\varrho$, where $\varrho=\int \varrho_{\bf p}\,{\rm d}^3{\bf
p}/(2\pi)^3$ and similar for antineutrinos. For the relevant values of
neutrino asymmetries this matter term dominates  and leads to synchronized
oscillations~\cite{Dolgov:2002ab, Wong:2002fa, Abazajian:2002qx}. The last
term in eq.\ \eqref{drhodt} corresponds to the effect of neutrino
collisions, i.e.\ interactions with exchange of momenta. Here we follow
the same considerations of ref.\ \cite{Pastor:2008ti}, where the reader
can find more details on the approximations made and related references.
In short, the collision terms for the off-diagonal components of
$\varrho_{\bf p}$ in the weak-interaction basis are momentum-dependent
damping factors, while collisions and pair processes for the diagonal
$\varrho_{\bf p}$ elements are implemented without approximations solving
numerically the collision integrals as in \cite{Mangano:2005cc}. These
last terms are crucial for modifying the neutrino distributions to achieve
equilibrium with $e^\pm$ and, indirectly, with photons.

We have solved numerically the EOMs for the matrices in flavor space of
neutrinos and antineutrinos with non-zero initial asymmetries. The
expansion of the universe is taken into account using comoving variables
as in \cite{Dolgov:2002ab}, where it was shown that flavor oscillations
between muon and tau neutrinos take place at $T_\gamma > 10$ MeV, when
interactions are very effective. For any initial values of the muon or tau
neutrino asymmetries, the combined effect of oscillations and collisions
is able to equilibrate the two flavors and hence leads to $\eta_{\nu_\mu}^{\rm in}=\eta_{\nu_\tau}^{\rm in}\equiv\eta_{\nu_x}^{\rm in}$.
Therefore, we start our numerical calculations for each case at $T=10$ MeV
and initial parameters $\eta_{\nu_e}^{\rm in}$ and the total asymmetry
$\eta_\nu=\eta_{\nu_e}^{\rm in}+2\,\eta_{\nu_x}^{\rm in}$. Note that hereafter by $\eta_\nu$ we denote the initial value of the total asymmetry, which is kept constant until the onset of $e^+-e^-$ annihilations, when it is diluted by the increase of the photon density (as in the case of the baryon asymmetry $\eta_b$).

All neutrino mixing parameters, except for $\theta_{13}$, are taken as the
best-fit values in \cite{Schwetz:2008er}. Modifying these parameters
within their allowed regions does not affect our results. Instead, the
value of $\theta_{13}$ plays an important role in the evolution of the
neutrino asymmetries \cite{Pastor:2008ti}. We thus consider either
$\theta_{13}=0$ or $\sin^2 \theta_{13}=0.04$, a value close to the upper
bound from neutrino experiments
\cite{Schwetz:2008er,Fogli:2008jx,GonzalezGarcia:2010er}.

\begin{figure}
\begin{center}
         \includegraphics[width=0.8\textwidth,angle=0]{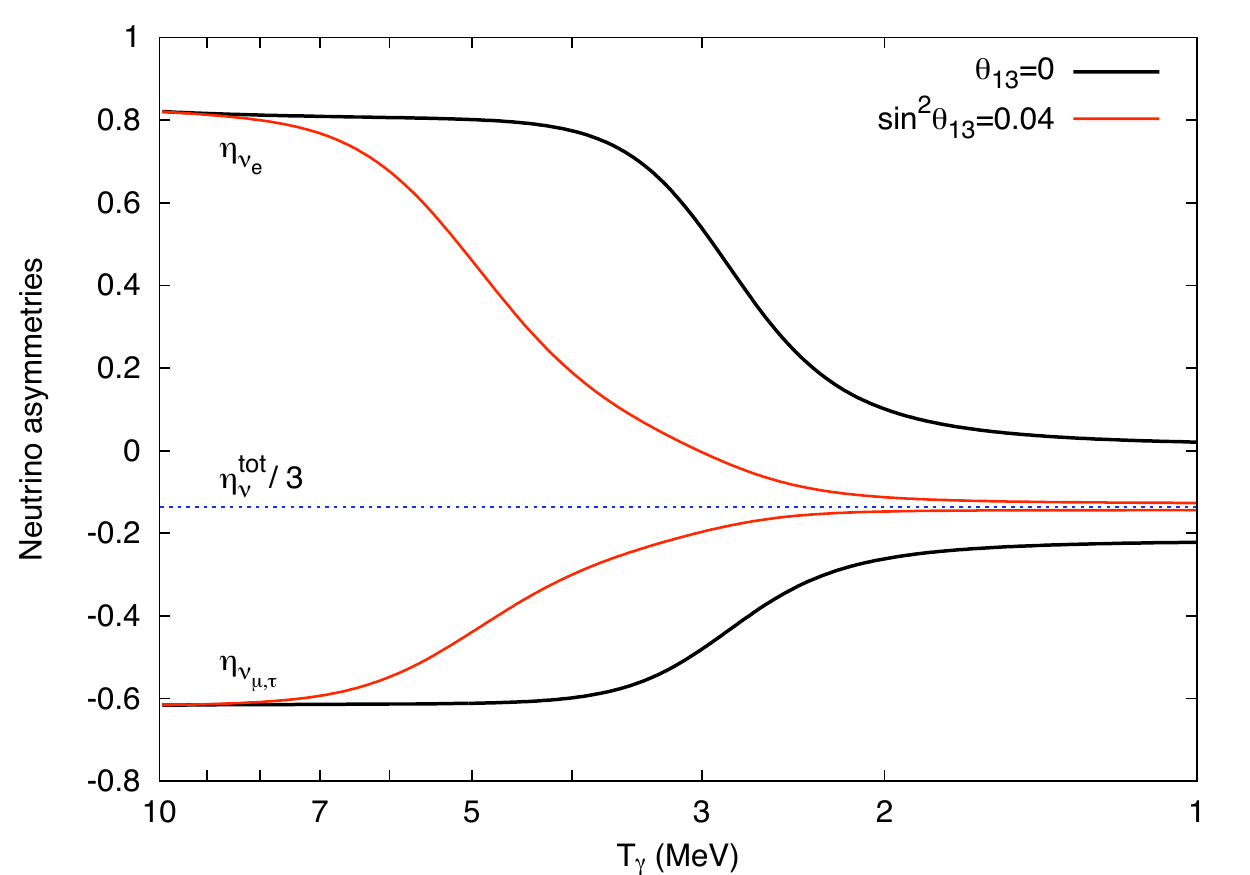}
    \end{center}
\caption{\label{evol_lnu} Evolution of the flavor neutrino asymmetries when
$\eta_\nu=-0.41$, and $\eta_{\nu_e}^{\rm in}=0.82$. The solid curves
correspond to vanishing $\theta_{13}$ (outer black lines) and
$\sin^2\theta_{13}=0.04$ (inner red lines). The total neutrino asymmetry
is constant and equal to three times the value shown (blue dotted line).}
\end{figure}
Let us describe the evolution of the flavor asymmetries with a specific
example. As shown in \cite{Pastor:2008ti}, large initial values of flavor
primordial asymmetries could satisfy the BBN bound only if
$\eta_{\nu_e}^{\rm in}$ and $\eta_{\nu_x}^{\rm in}$  have opposite signs
and the total asymmetry is not very different from zero. A case with
$\eta_\nu=0$ was the main example presented in \cite{Pastor:2008ti}. Here
we choose instead a benchmark case with non-zero negative total asymmetry,
$\eta_\nu=-0.41$, and $\eta_{\nu_e}^{\rm in}=0.82$. The evolution of the
flavor asymmetries is found from the numerical solution of the EOMs and is
shown in Figure \ref{evol_lnu}. Other choices of the initial asymmetries
will lead to different final values, but the overall behavior of the evolution is similar to 
the case shown here.

One can see in Figure \ref{evol_lnu} the effect of flavor oscillations on
the evolution of neutrino asymmetries with the universe temperature and
the dependence on the value of $\theta_{13}$. If this mixing angle is
close to the present experimental upper bound, flavor oscillations are
effective around $T_\gamma\sim 8$ MeV (inner red lines) when neutrinos are
still in good thermal contact with the ambient plasma.\footnote{Here we
consider only the case of normal hierarchy, $\Delta m^2_{\rm atm}>0$. If
we choose an inverted hierarchy the results are very similar, except that
equilibrium among the flavor asymmetries is reached slightly earlier.} The
neutrino distributions evolve keeping an equilibrium form, and the total
asymmetry is almost equally distributed among the three flavors. However,
the final value of the electron neutrino asymmetry is too different from
zero and this case is not allowed by BBN, as we will see in Section
\ref{sec:BBN}. Instead, for $\theta_{13}=0$ flavor oscillations begin only
at $T_\gamma\lesssim 3$ MeV, when weak interactions are not frequent
enough to keep the neutrino spectra in equilibrium. There is no full
equipartition of the total $\eta_\nu$ among the three flavors, although
collisions lead to final values of the flavor asymmetries closer to
$\eta_\nu/3$ than those expected in an MSW matter neutrino conversion. In
this case the final value of the electron neutrino asymmetry is close
enough to zero at the onset of BBN to lie in the favored region. If
$\sin^2\theta_{13}\lesssim 10^{-3}$, the outcome is very close to the case
of vanishing  $\theta_{13}$.

\begin{figure}
\begin{center}
         \includegraphics[width=0.8\textwidth,angle=0]{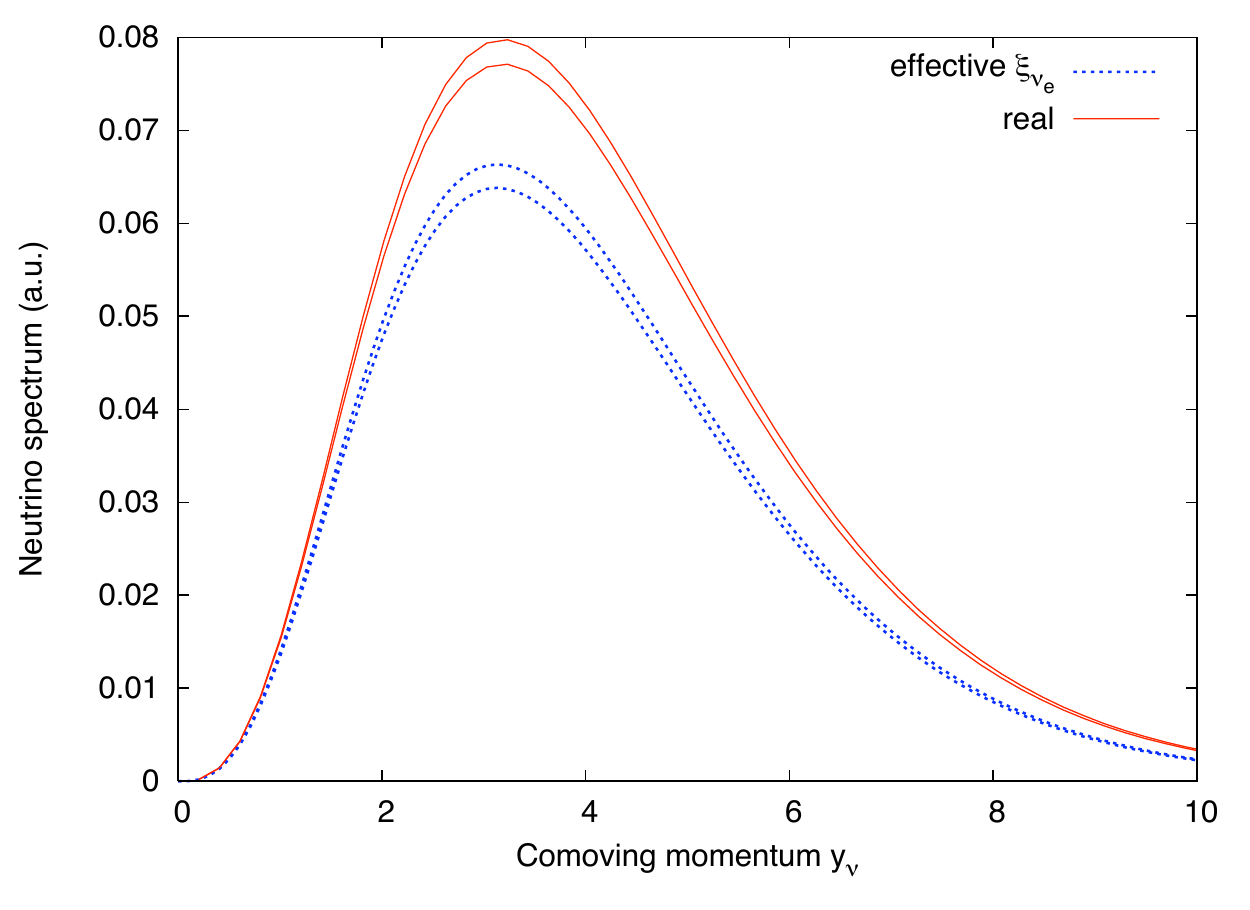}
    \end{center}
\caption{\label{drhofinal} The final energy spectra of relic electron  
neutrinos and antineutrinos in arbitrary units for the same case 
of Figure  \ref{evol_lnu} with vanishing $\theta_{13}$. Upper (lower) solid line stands for electron neutrino (antineutrino)  calculated numerically  (label ''real''). Upper (lower) dotted line stands for electron neutrino (antineutrino)  described by a Fermi/Dirac distribution just characterized by the same effective value of the electron neutrino  
degeneracy parameter.}
\end{figure}

In Figure \ref{drhofinal} we show the final energy spectra of relic electron  
neutrinos and antineutrinos in arbitrary units for the same case 
of Figure  \ref{evol_lnu} with vanishing $\theta_{13}$. The upper (lower) solid line stands for 
the spectra of electron neutrinos (antineutrinos)  calculated numerically, while the 
corresponding dotted lines
are described by a Fermi/Dirac distribution just characterized by the same effective value of the electron neutrino  degeneracy parameter  (as used in  all analyses before \cite{Pastor:2008ti}). Both cases lead to the  same value of the electron neutrino asymmetry but the real calculation shows  that an excess of radiation in neutrinos remains.

\begin{figure}
\begin{center}
         \includegraphics[width=0.8\textwidth,angle=0]{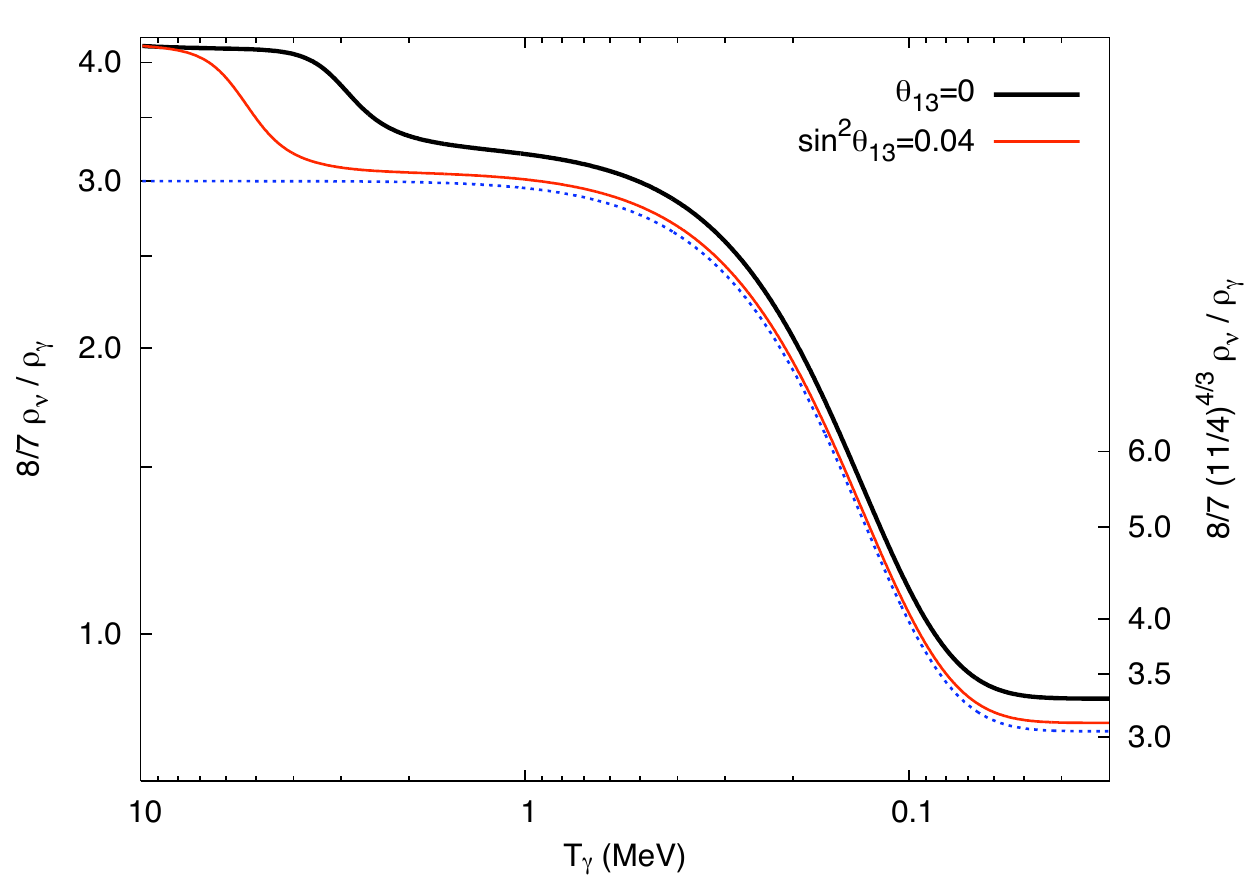}
    \end{center}
\caption{\label{evol_Neff} Evolution of the neutrino energy density for the
same case as in Figure \ref{evol_lnu}. The vertical axis is marked with
$N_{\rm eff}$, left before $e^+e^-$ annihilation, right afterwards. The
solid curves correspond to vanishing $\theta_{13}$ (upper black line) and
$\sin^2\theta_{13}=0.04$ (lower red line). The case without asymmetries is
shown for comparison (blue dotted line).}
\end{figure}

In Figure \ref{evol_Neff} we show the evolution of the ratio of neutrino
to photon energy densities, $\rho_\nu/\rho_\gamma$, properly normalized so
that it corresponds to $N_{\rm eff}$ at early and late times as in
\cite{Pastor:2008ti}. The fast drop of $\rho_\nu/\rho_\gamma$ at
$T\sim0.2$~MeV represents photon heating by $e^+e^-$ annihilations. The
case without asymmetries (dotted line) ends at late times at $N_{\rm
eff}=3.046$ instead of~3 because of residual neutrino
heating~\cite{Mangano:2005cc}. We also show (solid lines) the evolution
for our main example, where initially $N_{\rm eff}=4.16$ for our choice of
neutrino asymmetries. One can see that as soon as oscillations become
effective reducing the flavor asymmetries, the excess of entropy is
transferred from neutrinos to the electromagnetic plasma, cooling the
former and heating the latter, but this process is only very effective for
\emph{large} values of $\theta_{13}$. While the final $N_{\rm eff}$ is
$3.1$ for $\sin^2 \theta_{13}=0.04$, for negligible $\theta_{13}$ a
significant deviation from equilibrium survives and leads to a final
enhanced value of $N_{\rm eff}=3.3$.

%%%%%%%%%%%%%%%%%%%%%%%%%%%%%%%%%%%%%%%%%%%%%%%%%%%%%%%%%%%%%%%%%%%%%%
\section{Results from BBN: constraints on total and electron neutrino asymmetries}
\label{sec:BBN}
%%%%%%%%%%%%%%%%%%%%%%%%%%%%%%%%%%%%%%%%%%%%%%%%%%%%%%%%%%%%%%%%%%%%%%

As well known, BBN depends upon neutrino distribution functions in two
ways. First of all, electron neutrinos and antineutrinos enter directly in
the charge current weak processes which rule the neutron/proton chemical
equilibrium. A change in the effective temperature of the distribution can
shift the neutron/proton ratio freeze out  temperature and thus modifies the
primordial $^4$He abundance. Similarly, a non-zero $\nu_e-\bar{\nu}_e$
asymmetry also changes chemical equilibrium towards a larger or smaller
neutron fraction for negative or positive values of $\xi_{\nu_e}$,
respectively. Furthermore, lepton asymmetries in all flavors translate into a
positive extra contribution to the neutrino energy density, speeding up
the expansion rate given by the Hubble parameter.

\begin{figure}
\begin{center}
         \includegraphics[width=0.6\textwidth,angle=0]{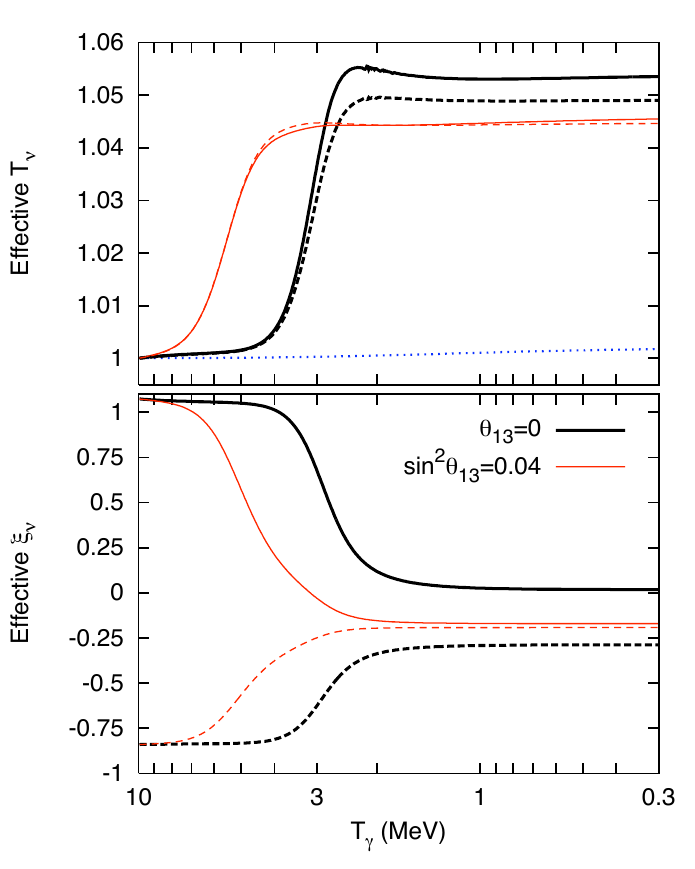}
    \end{center}
\caption{\label{evol_Txie} Evolution of the effective comoving temperatures
and degeneracy parameters of electron (solid lines) and muon or tau (dashed lines) 
neutrinos for the same case as in Figure \ref{evol_lnu}. Both the case of vanishing
$\theta_{13}$ (thick black lines) and  $\sin^2\theta_{13}=0.04$ (thin red
lines) are shown. The effective temperature for the case without  asymmetries is 
shown in the upper panel for comparison (blue dotted line).}
\end{figure}
Differently than in the approximated standard treatments, where both neutrino asymmetries
and the extra contribution to $N_{\rm{eff}}$ due to $\xi_{\nu_\alpha}$,
see eq. \eqref{deltan}, are considered as constant parameters, in the present
analysis we exactly follow the evolution of the neutrino distribution versus the
photon temperature $T_\gamma$, which is our evolution
parameter.\footnote{For another example of BBN calculations with
arbitrarily-specified, time-dependent neutrino and antineutrino
distribution functions, see Ref.\ \cite{Smith:2008ic}.} To this end we
have changed the public numerical code $\mathtt{PArthENoPE}$
\cite{Pisanti:2007hk,parthenope} as follows. For any given initial values
(at $T_\gamma= 10$ MeV) for the total neutrino asymmetry
$\eta_\nu=\sum_\alpha \eta_{\nu_\alpha}$, unchanged by flavor oscillations, and
electron neutrino asymmetry $\eta_{\nu_e}^{\rm in}$ we obtain, as
described in the previous section, the time dependent neutrino
distributions. The latter are then fitted in terms of Fermi-Dirac
functions with the two evolving parameters $T_{\nu_\alpha}(T_\gamma)$ and
$\xi_{\nu_{\alpha}}(T_\gamma)$. An example of their evolution is shown in Figure \ref{evol_Txie}
for the same choice of initial asymmetries as in the case described in
Section \ref{sec:dynamics}. Weak rates are then averaged over the
corresponding electron (anti)neutrino distribution. The Hubble parameter
is also modified to account for the actual evolution of total neutrino
energy density.

The final abundances of both the ratio $^2$H/H and the $^4$He mass
fraction, $Y_p$, are numerically computed as a function of the input
parameters $\eta_\nu$ and $\eta_{\nu_e}^{\rm in}$ and compared with the
corresponding experimental determinations. The baryon density parameter
has been set to the value determined by the 7-year WMAP result, $\Omega_b
h^2 = 0.02260 \pm 0.00053$ (68\% C.L.) \cite{Komatsu:2010fb}.\footnote{The
allowed region of $\Omega_b$ in extended cosmological models with free
$N_{\rm eff}$ does not differ significantly.} 

To get confidence intervals for  $\eta_\nu$ and
$\eta_{\nu_e}^{\rm in}$, one can construct the likelihood
function 
\be {\mathcal{L}}(\eta_\nu, \eta_{\nu_e}^{\rm in})\propto \exp\left(
-\chi^2(\eta_\nu, \eta_{\nu_e}^{\rm in})/2\right) \vv 
\ee 
with 
\be \chi^2(\eta_\nu, \eta_{\nu_e}^{\rm in}) = \sum_{ij} [
X_i(\eta_\nu, \eta_{\nu_e}^{\rm in}) - X_i^{obs} ] W_{ij}(\eta_\nu, \eta_{\nu_e}^{\rm in}) [ X_j(\eta_\nu, \eta_{\nu_e}^{\rm in}) - X_j^{obs} ] \pp
\label{eq:chi2}
\ee 
The proportionality constant can be obtained by requiring normalization to unity, and $W_{ij}(\eta_\nu, \eta_{\nu_e}^{\rm in})$ denotes the inverse covariance matrix \cite{serpico}, 
\be W_{ij}(\eta_\nu, \eta_{\nu_e}^{\rm in}) = [ \sigma_{ij}^2 +
\sigma_{i,exp}^2 \delta_{ij} + \sigma_{ij,other}^2 ]^{-1} \vv \ee
where $\sigma_{ij}$ and $\sigma_{i,exp}$ represent the nuclear
rate uncertainties and experimental uncertainties of nuclide
abundance $X_i$, respectively \cite{serpico}, while by
$\sigma_{ij,other}^2$ we denote the propagated squared error
matrix due to all other input parameter uncertainties ($\tau_n$,
$G_{\rm N}$, $\Omega_b h^2$, \ldots). In our case we consider in eq.\ \eqref{eq:chi2} as $X_i$ the quantities  
 $^2$H/H and $Y_p$ only.

Let us now briefly discuss the set of data we have used in our study. The
$^2$H/H number density is obtained by averaging seven determinations
obtained in different Quasar Absorption Systems, as in \cite{Iocco:2008va}
\be
^2{\rm H/H}= (2.87 \pm 0.22) \times 10^{-5} \vv \label{dataD}
\ee
where the quadratic error has been enlarged by the value of the reduced
$\chi^2$ to account for the dispersion of measurements for this dataset,
$\chi^2_{\rm min}/6=3.6$, see \cite{Iocco:2008va}.

For the $^4$He mass fraction we consider two different determinations. One
is the result of the data collection analysis performed in
\cite{Iocco:2008va},
\be
Y_p=0.250 \pm 0.003\, . 
\label{he1}
\ee
More recently, new studies of metal poor H II regions have appeared in the
literature \cite{Izotov:2010ca,Aver:2010wq,Aver:2010wd}. While these groups both agree
on a larger central value with respect to the result of eq. \eqref{he1},
which incidentally seems to pin down a value for $N_{\rm{eff}}>3$ at
$2\sigma$, a different estimate of possible systematic
effects which dominate the total uncertainty budget is quoted in \cite{Izotov:2010ca} and \cite{Aver:2010wq}, with
\cite{Aver:2010wq} quoting a larger error, of the order of 4\%. 
\bea
Y_p & = & 0.2565 \pm 0.0010 {\rm (stat.)} \pm 0.0050 {\rm (syst.)}  \,\,\,\,\, 
\mbox{\cite{Izotov:2010ca}} \vv \\
Y_p & = &  0.2561 \pm 0.0108  \,\,\,\,\, \mbox{\cite{Aver:2010wq}} \pp
\eea
Finally, in a recent paper \cite{Aver:2010wd}, Markov Chain Monte Carlo method was exploited  to 
determine the $^4$He  abundance, and the uncertainties derived from observations of metal poor nebulae 
finding
\be
Y_p = 0.2573 \pm 0.0033 \, . \label{he2}
\ee
In the following, we will use the two results of eq.s \eqref{he1} and \eqref{he2}. While their uncertainties are the same, they differ for the central value, actually the smaller and higher of all results reported above, a fact which will produce two different bounds on the electron neutrino asymmetry. 
\begin{figure}[t]
\begin{center}
         \includegraphics[width=0.49\textwidth,angle=0]{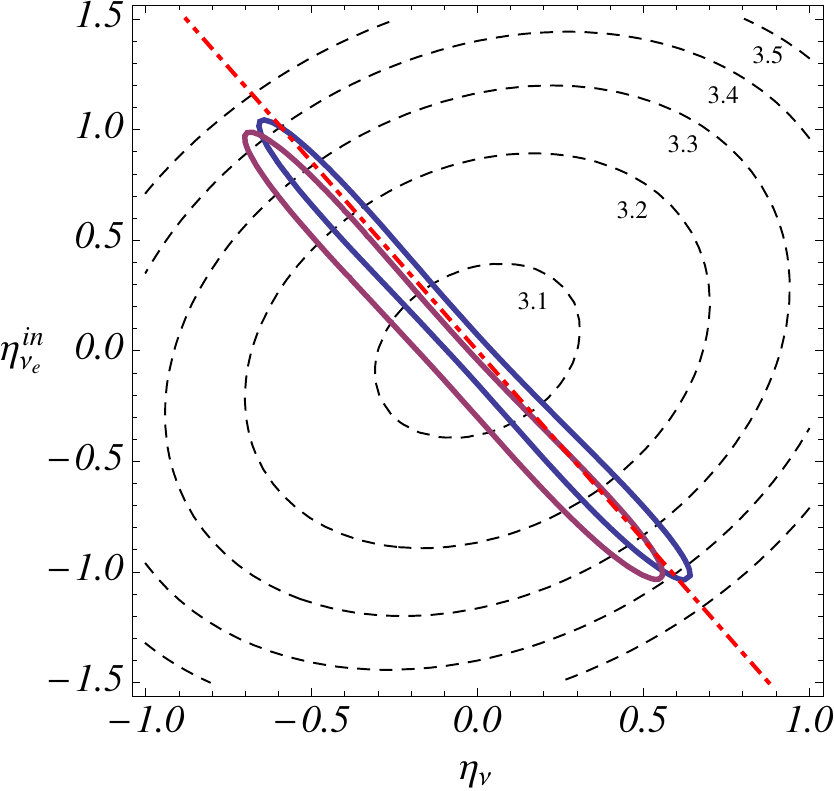}
     \includegraphics[width=0.49\textwidth,angle=0]{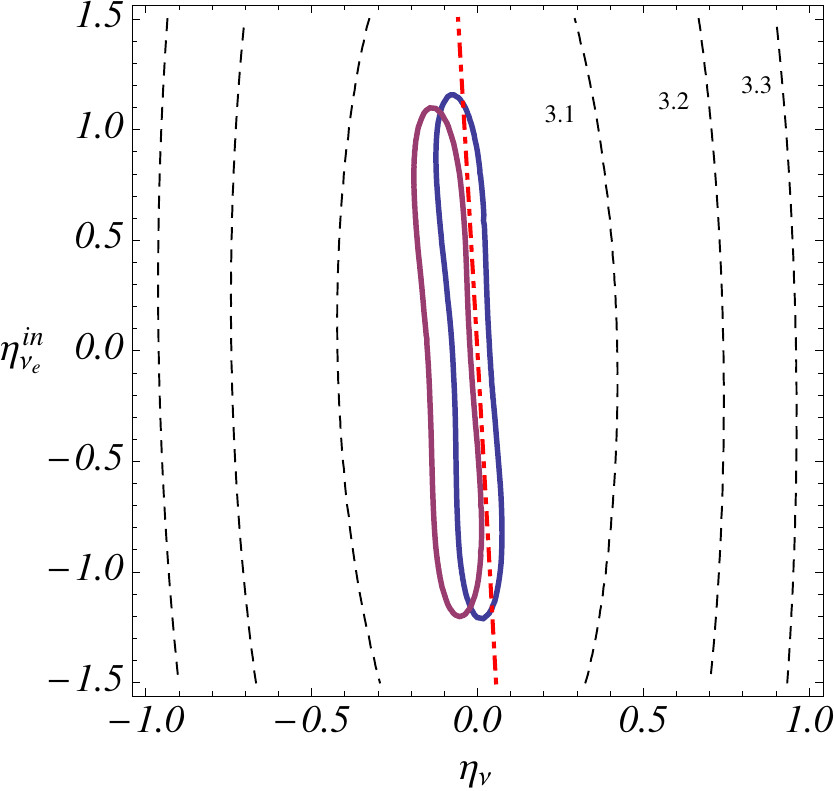}
    \end{center}
\caption{\label{bounds_eta} The 95\% C.L.\ contours from our BBN analysis
in the $\eta_\nu-\eta_{\nu_e}^{\rm in}$ plane for $\theta_{13}=0$ (left)
and $\sin^2 \theta_{13}=0.04$ (right). The two contours correspond to the
different choices for the primordial $^4$He abundances of eqs.\
\eqref{he1} (blue) and \eqref{he2} (purple). The (red)
dot-dashed line is the set of  values of $\eta_\nu$ and $\eta_{\nu_e}^{\rm
in}$ which, due to flavor oscillations, evolve towards a vanishing final
value of electron neutrino asymmetry $\eta_{\nu_e}^{\rm fin}$. We also
report as dashed lines the iso-contours for different values of $N_{\rm
eff}$, the effective number of neutrinos after $e^+e^-$ annihilation
stage.}
\end{figure}

\begin{figure}%[b]
	\begin{center}
     \includegraphics[width=0.49\textwidth,angle=0]{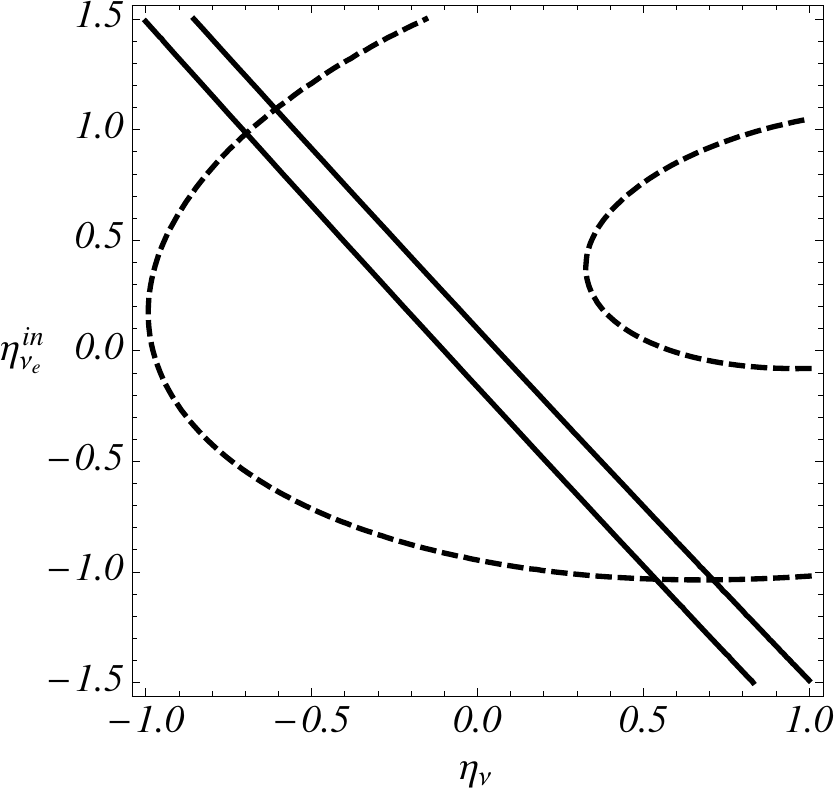}
     \end{center}
\caption{\label{He4_D} Bounds in the $\eta_{\nu_e}^{\rm in}$ - $\eta_\nu$ plane for each nuclear yield. Areas between the lines correspond to 95\% C.L. regions singled out by the $^4$He mass fraction (solid lines) of eq.\ \eqref{he1},  and Deuterium (dashed lines) as in eq.\ \eqref{dataD}. }
\end{figure}
\begin{figure}[t]
\begin{center}
         \includegraphics[width=0.49\textwidth,angle=0]{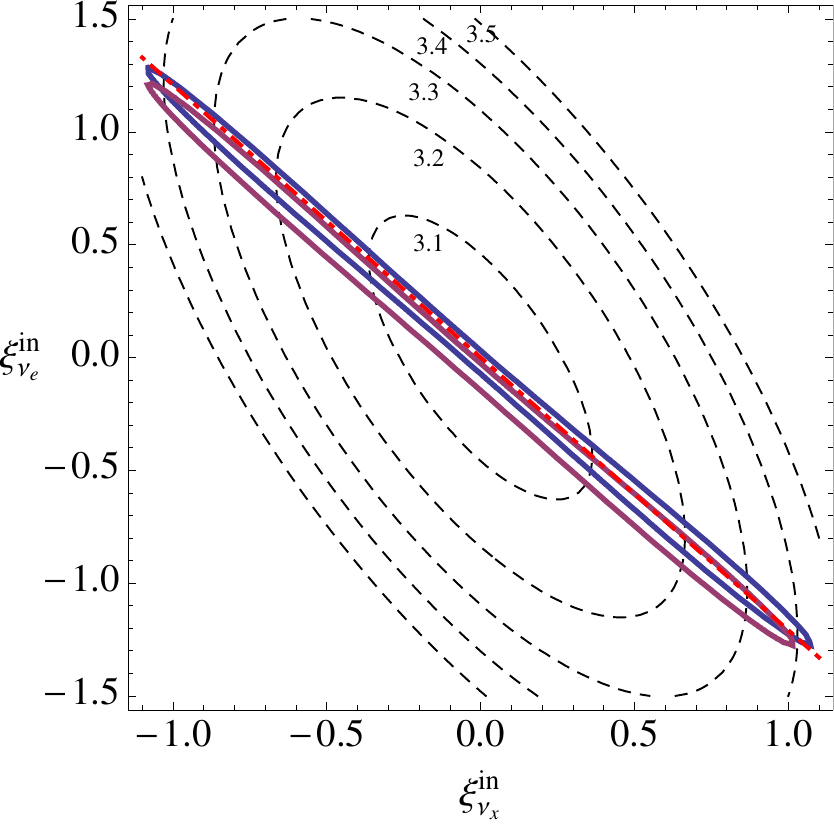}
     \includegraphics[width=0.49\textwidth,angle=0]{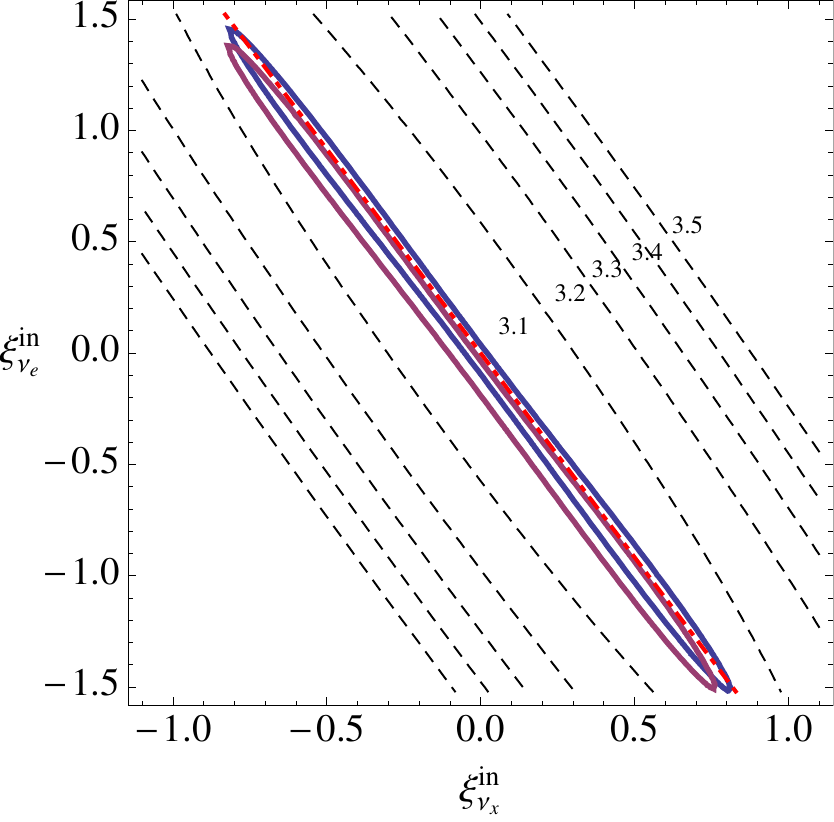}
    \end{center}
\caption{\label{bounds_xi} Same results as in Figure \ref{bounds_eta} in the plane of initial flavor
degeneracy parameters $\xi_{\nu_x}^{\rm in}$ and $\xi_{\nu_e}^{\rm in}$.}
\end{figure}

\begin{figure}%[b]
\begin{center}
     \includegraphics[width=0.49\textwidth,angle=0]{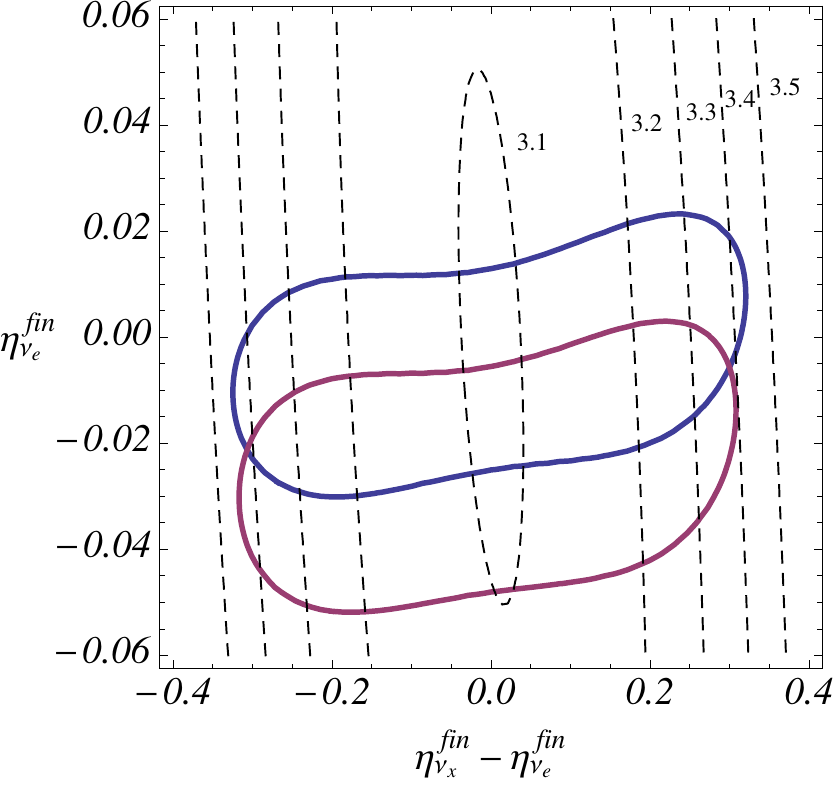}% &
     \includegraphics[width=0.49\textwidth,angle=0]{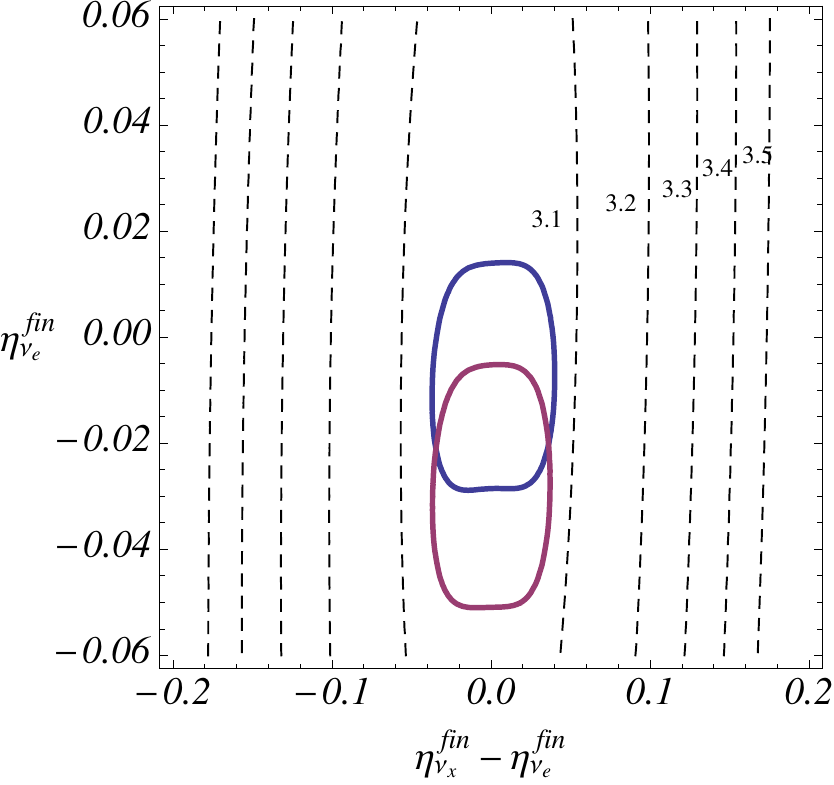}
    \end{center}
\caption{\label{bounds_diff} Same results as in Figure \ref{bounds_eta}
in the plane of $\eta_{\nu_e}^{\rm fin}$ and the difference
$\eta_{\nu_x}^{\rm fin}-\eta_{\nu_x}^{\rm fin}$, where the superscript
indicates that the asymmetries are evaluated at the onset of BBN,
$T_\gamma \sim 1$ MeV, after flavor oscillations shuffled the initial
$\eta_{\nu_\alpha}^{\rm in}$.}
\end{figure}

The 95\% C.L.\ contours for the total asymmetry $\eta_\nu$ and the initial
value of the electron neutrino parameter  $\eta_{\nu_e}^{\rm in}$ are
shown in Figure \ref{bounds_eta} for the adopted determinations of $^2$H and
$^4$He and for two different choices of $\theta_{13}$. In both cases the
contours are close to and aligned along the red dot-dashed line which represents the set of
initial values for the asymmetries which eventually evolve toward a
vanishing final electron neutrino asymmetry, $\eta_{\nu_e}^{\rm
fin}\simeq0$, which is preferred by $^4$He data. We recall that $^4$He is
strongly changed if neutron/proton chemical equilibrium is shifted by a
large value of $\nu_e-\bar{\nu}_e$ asymmetry around the freezing of weak
rates ($T_\gamma \sim 0.8$ MeV). For large $\theta_{13}$, oscillations
efficiently mix all neutrino flavors and at BBN $\eta_{\nu_\alpha} \sim
\eta_\nu/3$, so the bound on $\eta_\nu$ is quite stringent, $-0.1 \lesssim
\eta_\nu \lesssim 0.1$, if we adopt the value of eq.\ \eqref{he1} for $Y_p$.
Instead, for the choice of eq.\ \eqref{he2} we find $-0.20 \lesssim \eta_\nu
\lesssim 0$,  i.e. a larger value for $Y_p$ singles out slightly negative values for $\eta_\nu$ (and $\eta_{\nu_e}^{\rm
fin}$), since the theoretical prediction for $Y_p$ grows in this case as the neutron-proton chemical equilibrium shift towards a larger neutron fraction at freeze-out . On the other
hand, for a vanishing $\theta_{13}$ the contours for  $\eta_\nu$ and
$\eta_{\nu_e}^{\rm in}$ show a clear anticorrelation, and even values of
order unity for both parameters are still compatible with BBN. The allowed
regions of the total neutrino asymmetry are summarized in
Table \ref{table_etanu}. 

We stress that for any value of $\theta_{13}$, the data on primordial deuterium, eq.\
\eqref{dataD}, is crucial for closing the allowed region that the $^4$He
bound fixes along the $\eta_{\nu_e}^{\rm fin}\simeq0$ line. In fact, though $^2$H is less sensitive than $^4$He to neutrino asymmetries and effective temperature which enter the Universe expansion rate, see e.g. \cite{Iocco:2008va}, yet including it in the analysis breaks the degeneracy between 
$\eta_{\nu_e}^{\rm in}$ and $\eta_\nu$ which is present when only $^4$He is used. This can be read from Figure \ref{He4_D} where the 95\% C.L. in the $\eta_{\nu_e}^{\rm in}$ - $\eta_\nu$ plane are shown for $^4$He and $^2$H separately,  for the case $\theta_{13}=0$. The solid lines bound the region of the plane compatible with the $^4$He measurement as in eq.\ \eqref{he1}, whereas the dashed contours correspond to Deuterium observation, see eq.\ \eqref{dataD}. The different shape of these two regions is due to the different dependence of nuclide abundances on  $\eta_{\nu_e}^{\rm in}$  and $\eta_\nu$, thus their combination breaks the degeneracy and leads to a close contour as shown in Figure \ref{bounds_eta}.

\begin{table}
\begin{center}
\begin{tabular}{l|c|c|}
{} & $\theta_{13}=0$ & $\sin^2 \theta_{13}= 0.04$\\
\hline
$Y_p=0.250 \pm 0.003$ &
$-0.66 < \eta_\nu < 0.63$ &
$-0.13 < \eta_\nu < 0.07$ \\
\hline
$Y_p = 0.2573 \pm 0.0033$ &
$-0.71 < \eta_\nu < 0.56$ &
$-0.20 < \eta_\nu < 0.02$ \\
\hline
\end{tabular}
\end{center}
\caption{\label{table_etanu} BBN bounds on the initial total neutrino asymmetry at 95\% C.L.}
\end{table}

It is also interesting to report our results in terms of other variables,
as in Figures \ref{bounds_xi}  and \ref{bounds_diff}. In the first case,
the BBN contours are shown in the plane of initial flavor degeneracy
parameters $\xi_{\nu_e}^{\rm in}$ and $\xi_{\nu_x}^{\rm in}$ while, in
Figure \ref{bounds_diff}, we consider a new pair of variables: the
electron neutrino asymmetry at the onset of BBN $\eta_{\nu_e}^{\rm fin}$,
and the difference $\eta_{\nu_x}^{\rm fin}-\eta_{\nu_e}^{\rm fin}$, which
in the standard analysis is usually assumed to be vanishing. One can see
from this figure that, while the (95\% C.L.) bound on $\eta_{\nu_e}^{\rm fin}$ is
independent of the value of $\theta_{13}$
\bea
&& -0.03 \leq \eta_{\nu_e}^{\rm fin} \leq 0.02\,  , \,\,\,\, {\rm for}~Y_p = 0.250 \pm 0.003 \vv \\
&& -0.05 \leq \eta_{\nu_e}^{\rm fin} \leq 0.003 \, , \,\,\,\, {\rm for}~Y_p = 0.2573 \pm 0.0033 \vv
\eea
the difference between the final $\nu_e$ and $\nu_x$ asymmetries strongly
depends upon this yet unknown mixing angle, as expected. In fact, for
large $\theta_{13}$ we recover the standard result, $\eta_{\nu_x}^{\rm
fin} \sim \eta_{\nu_e}^{\rm fin}$, due to efficient mixing by oscillations
and collisions, while for $\theta_{13}=0$ the two asymmetries can be
different. The following ranges are in good agreement with BBN data at 95 \% C.L.,
\emph{independently} of the adopted value for $Y_p$
\bea
 -0.3 \leq \eta_{\nu_x}^{\rm fin} - \eta_{\nu_e}^{\rm fin} \leq 0.3\,  , && \,\,\,\, \sin^2 \theta_{13}= 0 \vv \\
 -0.04 \leq \eta_{\nu_x}^{\rm fin} - \eta_{\nu_e}^{\rm fin} \leq 0.04 \, , && \,\,\,\, \sin^2 \theta_{13}= 0.04  \pp
\eea
We conclude that, in particular for $\theta_{13}= 0$ oscillations lead to quite different neutrino asymmetries in $e$ and $\mu/\tau$ flavors, still being in good agreement with BBN, differently than what was previously assumed in the literature, see \cite{Lunardini:2000fy,Dolgov:2002ab,Wong:2002fa,Abazajian:2002qx}, and \cite{Iocco:2008va,Hansen:2001hi,Barger:2003zg,Barger:2003rt,Cuoco:2003cu,Cyburt:2004yc,
Serpico:2005bc,Simha:2008mt,Krauss:2010xg} for BBN analyses.

\begin{figure}%[b]
	\begin{center}
     \includegraphics[width=0.8\textwidth,angle=0]{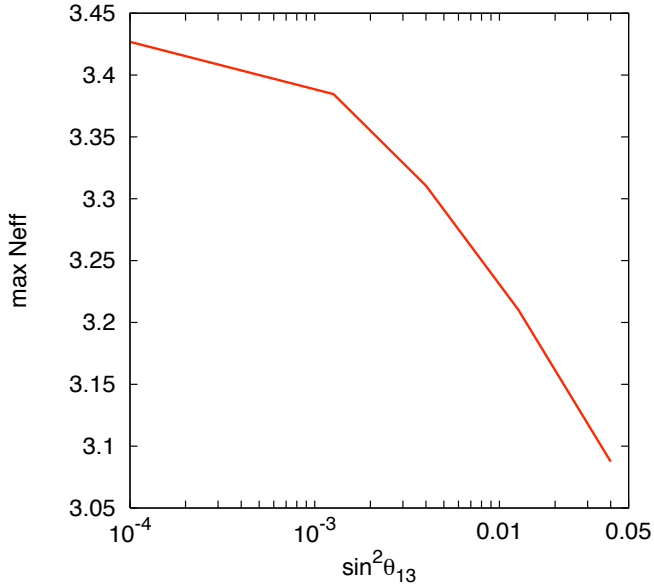}
     \end{center}
\caption{\label{theta13_Neff} Largest values of $N_{\rm{eff}}$ from primordial neutrino
asymmetries compatible with BBN, as a function of $\theta_{13}$.}
\end{figure}

In Figures \ref{bounds_eta}-\ref{bounds_diff} we also plot iso-contours
for the value of the effective number of neutrinos, $N_{\rm{eff}}$,
evaluated after $e^+e^-$ annihilations. For large $\theta_{13}$ BBN data
bound $N_{\rm{eff}}$ to be very close to the standard value 3.046, since
all asymmetries should be very small in this case and flavor oscillations
modify the neutrino distributions while neutrinos are still strongly
coupled to the electromagnetic bath. Therefore, we do not expect
non-thermal features in the neutrino spectra in this case, since
scatterings and pair processes allow for an efficient transfer of any
entropy excess. On the other hand, for vanishing $\theta_{13}$, larger
values of $N_{\rm{eff}}$ are still compatible with BBN data, up to values
of the order of 3.4 at 95\% C.L. The dependence of the largest achievable value of $N_{\rm{eff}}$
on the value of the mixing angle $\theta_{13}$, obtained by spanning in the asymmetry parameter plane the region compatible with BBN, is reported in Figure \ref{theta13_Neff}. 

It is worth noticing that the final values of $N_{\rm eff}$,
in particular for large final asymmetries of $\nu_x$, are also slightly
larger than the $N_{\rm eff}$ that one would obtain using the equilibrium
expression of eq.\ \eqref{eta-eq}. For example, if we take
$\eta_{\nu_e}^{\rm fin}=0$ and $\eta_{\nu_x}^{\rm fin}=0.3$, a point on
the boundary of the BBN contours (see Figure \ref{bounds_diff}) and
compute the corresponding effective chemical potentials, using eq.\
\eqref{deltan} one gets $N_{\rm{eff}} = 3.2$, while the actual value is
larger, $N_{\rm{eff}} = 3.4$, a signal that in this case the interplay of
solar-like oscillations and neutrino freeze-out has produced indeed, a
mild non-thermal distortion in neutrino distributions.

Finally, we notice that our BBN results correspond to a minimal scenario
with primordial neutrino asymmetries, as we do not consider a possible
extra contribution to $N_{\rm{eff}}$ coming from relativistic degrees of
freedom other than standard active neutrinos. Their effect is known to
produce looser bounds on neutrino asymmetries, as they speed up expansion
and thus, can compensate the effect of a positive $\nu_e-\bar{\nu}_e$
asymmetry. We have explicitly checked that, for some choices of primordial
asymmetries, the addition of extra radiation does not modify the evolution
of flavor neutrino asymmetries. Of course, in such a case the contribution
to the energy density of the additional relativistic degrees of freedom
adds up to the surviving excess to $N_{\rm{eff}}$ arising from neutrino
asymmetries.

%%%%%%%%%%%%%%%%%%%%%%%%%%%%%%%%%%%%%%%%%%%%%%%%%%%%%%%%%%%%%%%%%%%%%%
\section{Conclusions}\label{sec:conclusions}
%%%%%%%%%%%%%%%%%%%%%%%%%%%%%%%%%%%%%%%%%%%%%%%%%%%%%%%%%%%%%%%%%%%%%%

In this paper we have calculated the evolution of neutrinos in the early
universe with initial flavor asymmetries, taking into account the combined
effect of collisions and oscillations. Our numerical results for the
neutrino momentum distributions, which can develop non-thermal features,
were used to find the primordial production of light elements, employing a
modified version of the $\mathtt{PArthENoPE}$ BBN code. Comparing with the recent
data on primordial $^2$H and $^4$He abundances, we have found the allowed
ranges for both the total asymmetry and the initial asymmetry in
electronic flavor. These BBN limits mostly depend on the value of
$\theta_{13}$, the only unknown mixing angle of neutrinos. As can be seen from Figure \ref{theta13_Neff}, for $\sin^2\theta_{13} \lesssim 10^{-3}$ this implies an
effective number of neutrinos bound to $N_{\rm eff}\lesssim 3.4$, independently of which of the experimental $^4$He mass fraction of eq.s \eqref{he1} and \eqref{he2} is adopted, whereas for larger values of $\theta_{13}$ the effective number of neutrinos is closely bound to the standard value 3.046 obtained for vanishing asymmetries.

In the near future it will be possible to improve our BBN constraints on
the lepton number of the universe. On one hand, thanks to the better
sensitivity of new neutrino experiments, either long-baseline or reactor,
we expect to have a very stringent bound on $\theta_{13}$ or eventually its
measurement (see e.g.\ \cite{Huber:2004ug}). Such results would lead to a
more restrictive BBN analysis on primordial asymmetries. On the other
hand,  in the next couple of years data on the anisotropies of the cosmic
microwave background from the Planck satellite \cite{Planck} will largely
reduce the allowed range of $N_{\rm eff}$, since the forecast sensitivity
is of the order of $0.4$ at $2\sigma$
\cite{Bowen:2001in,Bashinsky:2003tk,Hamann:2010pw}.
Indeed, suppose that Planck data confirm a value of  $N_{\rm eff}> 3$.
An excess of radiation up to $0.4-0.5$ could imply the presence
of a significative degeneracy in the neutrino sector, but this would depend  on the result of $\theta_{13}$
measurement. A vanishing $\theta_{13}$ would allow such possibility, but a measured $\theta_{13}$ in the next generation of experiments {\it would imply} the presence of extra degrees of freedom others than active neutrinos. In case of a  much larger result by Planck, namely $N_{\rm eff}  > 4$,
it would be impossible to explain such a result in terms of
primordial neutrino asymmetries only, and alternative cosmological
scenarios with additional relativistic species, such as sterile neutrinos
(see for instance \cite{Melchiorri:2008gq}) would be strongly favored.

%%%%%%%%%%%%%%%%%%%%%%%%%%%%%%%%%%%%%%%%%%%%%%%%%%%%%%%%%%%%%%%%%%%%%%
\acknowledgments
%%%%%%%%%%%%%%%%%%%%%%%%%%%%%%%%%%%%%%%%%%%%%%%%%%%%%%%%%%%%%%%%%%%%%%

We would like to thank Georg Raffelt for his useful comments on an earlier version of this paper.
G.\ Mangano, G.\ Miele, O.\ Pisanti and S.\ Sarikas acknowledge support by
the {\it Istituto Nazionale di Fisica Nucleare} I.S. FA51 and the PRIN
2010 ``Fisica Astroparticellare: Neutrini ed Universo Primordiale'' of the
Italian {\it Ministero dell'Istruzione, Universit\`a e Ricerca}. S.\
Pastor was supported by the Spanish grants FPA2008-00319 and Multidark CSD2009-00064
(MICINN) and PROMETEO/2009/091 (Generalitat Valenciana), and by the EC
contract UNILHC PITN-GA-2009-237920. This research was also supported by a
Spanish-Italian MICINN-INFN agreement, refs.\ FPA2008-03573-E and
ACI2009-1051.

\end{document}